\documentclass[a4paper,fleqn]{cas-dc}
\usepackage[numbers]{natbib}
\usepackage{amsthm}
\usepackage{amssymb}
\usepackage{physics}
\usepackage{braket}
\usepackage{listings}
\usepackage{diagbox}
\usepackage{breakurl}
\usepackage{hyperref}
\usepackage{amsfonts}
\usepackage{graphicx}
\usepackage{float}
\usepackage[T1]{fontenc}

\usepackage{jlcode}
\def\code#1{{\texttt{#1}}}

\def\code#1{{\texttt{#1}}}

\begin{document}
\let\WriteBookmarks\relax
\def\floatpagepagefraction{1}
\def\textpagefraction{.001}

\shorttitle{QuanEstimation.jl}

\shortauthors{H.-M. Yu and J. Liu}

\title [mode = title]{QuanEstimation.jl: An open-source Julia framework for quantum parameter estimation}                      

\author[1,2]{Huai-Ming Yu}

\author[1,2]{Jing Liu}[orcid=0000-0001-9944-4493]
\cormark[1]
\ead{liujing@hainanu.edu.cn}

\affiliation[1]{organization={Center for Theoretical Physics and School of Physics and Optoelectronic Engineering, Hainan University},
    city={Haikou},
    postcode={570228}, 
    state={Hainan},
    country={China}}
    
\affiliation[2]{organization={School of Physics, Huazhong University of Science and Technology},
    city={Wuhan},
    postcode={430074}, 
    state={Hubei},
    country={China}}    
   
\cortext[cor1]{Corresponding author}

\begin{abstract}
As the main theoretical support of quantum metrology, quantum parameter estimation 
must follow the steps of quantum metrology towards the applied science and industry. 
Hence, optimal scheme design will soon be a crucial and core task for quantum parameter 
estimation. To efficiently accomplish this task, software packages aimed at computer-aided 
design are in high demand. In response to this need, we hereby introduce QuanEstimation.jl, 
an open-source Julia framework for scheme evaluation and design in quantum parameter estimation. 
It can be used either as an independent package or as the computational core of the recently 
developed hybrid-language (Python-Julia) package QuanEstimation [Phys. Rev. Res. 4 (4) (2022) 043057]. 
Utilizing this framework, the scheme evaluation and design in quantum parameter estimation can be 
readily performed, especially when quantum noises exist. 
\end{abstract}



\maketitle

\section{Introduction}

Quantum technologies have encountered a fast-developing era in recent years, and are now being 
enthusiastically pursued by international technology companies and governments worldwide. Together 
with the artificial intelligence, quantum technologies have been treated as the major origins of 
the next-generation technologies, and even the next industrial revolution. 

Quantum technologies are founded on the principles of quantum mechanics and use quantum systems or 
quantum features to achieve advantages that classical systems cannot realize. As a core aspect 
of quantum technologies, quantum metrology utilizes quantum systems to perform precise measurements of 
physical parameters, such as the strength and frequency of an electromagnetic field or a signal. Its 
value has been successfully proved by many remarkable examples, such as the optical clocks~\cite{Beloy2021} 
and atomic magnetometers~\cite{Bao2020}. 

Quantum parameter estimation~\cite{Helstrom1976,Holevo1982} is the major theoretical support of quantum 
metrology due to the statistical nature of quantum systems. After decades of development, many elegant 
mathematical tools and optimal schemes have been provided and studied for quantum parameter estimation 
in various measurement scenarios. In practice, quantum noises are inevitable in the process of quantum 
parameter estimation, which usually affects the optimality of the optimal schemes given in noiseless 
scenarios, and different physical systems may face different dominant quantum noises. These facts 
indicate that the scheme design in the presence of noise usually needs to be performed case by case. 
Therefore, software for scheme design is a natural requirement in the industrialization process of quantum 
parameter estimation and quantum metrology. 

\begin{figure}[bp]
\centering\includegraphics[width=8.cm]{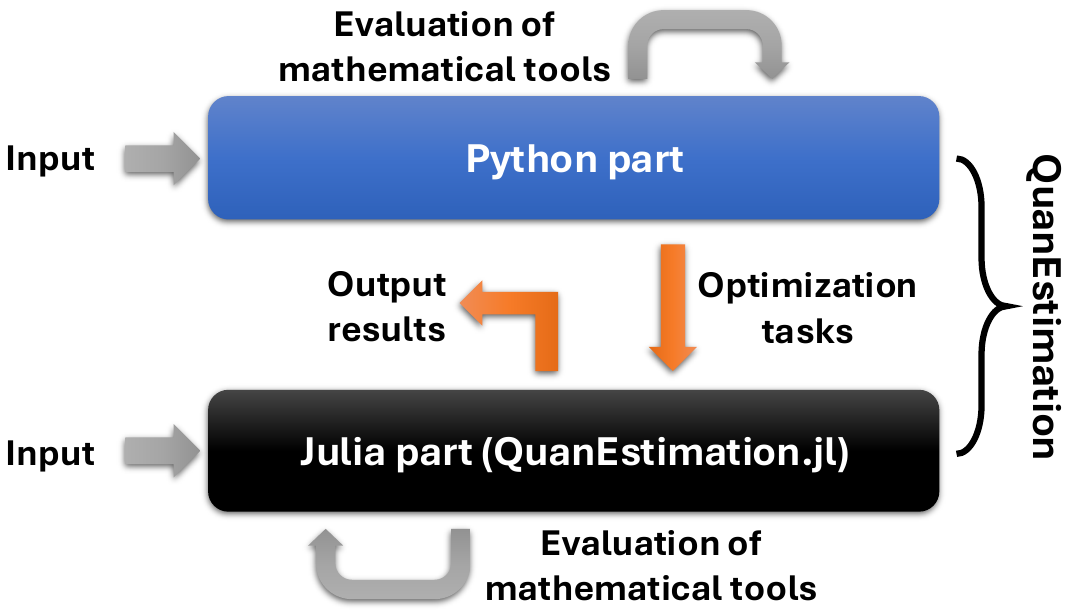}
\caption{Logical relation between the Python part and Julia part (QuanEstimation.jl) of QuanEstimation.   
\label{fig:relation}}
\end{figure}

\begin{figure*}[tp]
\centering\includegraphics[width=17.5cm]{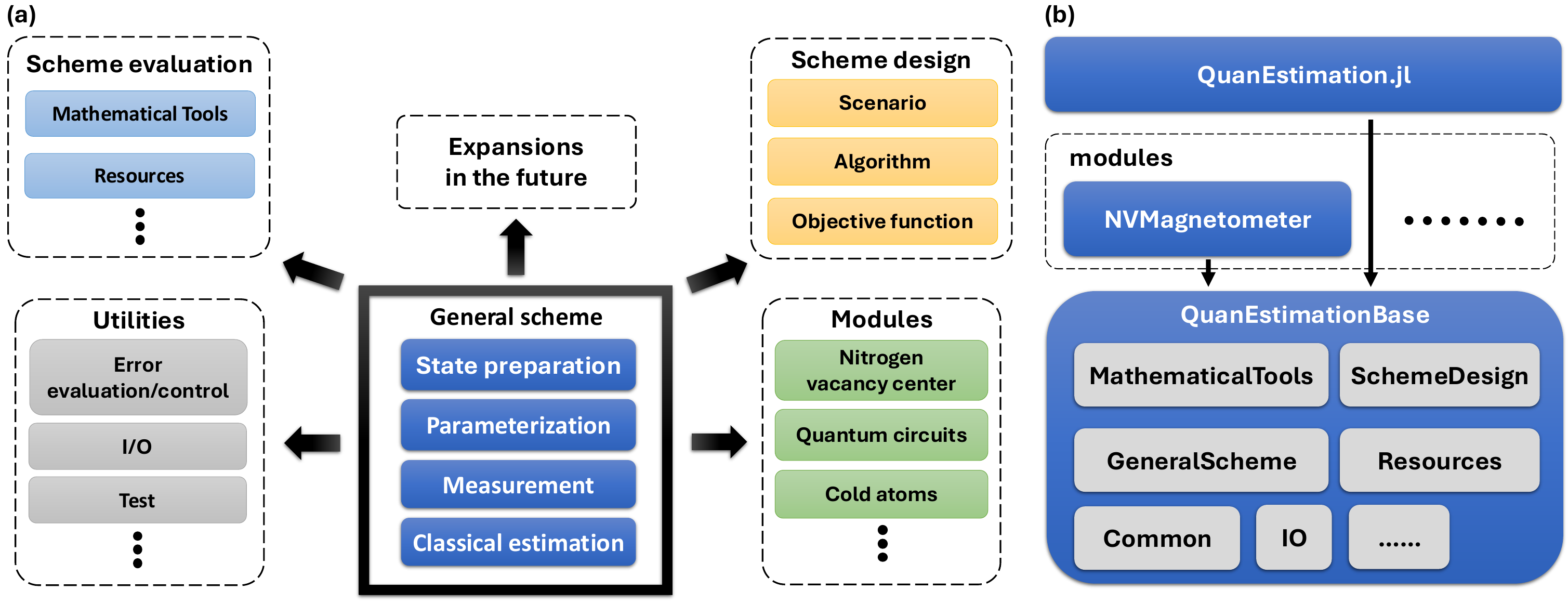}
\caption{(a) The architecture and (b) package structure of QuanEstimation.jl. 
\label{fig:arch}}
\end{figure*}

Due to this requirement, in 2022 we announced an open-source hybrid-language (Python-Julia) package named 
QuanEstimation for scheme evaluation and design in quantum parameter estimation~\cite{Zhang2022}. Python is 
the interface of the package due to its popularity in the scientific community, and the computational core is written 
in Julia since it provides superior numerical efficiency in the scheme design. In recent years, Julia~\cite{Bezanson2012} 
has become an emerging platform for computation packages in quantum information, and several very popular and useful 
packages have been developed based on it, including QuantumToolbox.jl~\cite{QTjl}, QuantumOptics.jl~\cite{Kramer2018}, 
QuantumInformation.jl~\cite{Gawron2018}, Yao.jl~\cite{Yaojl}, and QuantumControl.jl~\cite{QCjl}. The Julia part of 
QuanEstimation is actually an independent and complete package named QuanEstimation.jl. In the hybrid-language 
package, the evaluation of various mathematical tools can be directly executed in the Python part. However, the 
optimization tasks in the scheme design will be transferred to the Julia part, namely QuanEstimation.jl,  and be 
executed, as illustrated in  Fig.~\ref{fig:relation}. Once the calculation is executed in QuanEstimation.jl, it will stay in 
this part until all results are obtained and output. In the meantime, as an independent and complete package, all the 
functions in the Python part also exist in QuanEstimation.jl. After the announcement of QuanEstimation, QuanEstimation.jl 
has been constantly updated and optimized structurally. Now it not only contains all the functions in the hybrid-language 
package, but also includes some new features like modules for specific physical systems.  

In this paper, the architecture and package structure of QuanEstimation.jl will be thoroughly introduced. 
Furthermore, the methods to set up a scheme, including the general method and specific system modules, and the 
process of scheme design will be discussed, and corresponding demonstration codes will be presented. The usage 
of specific system modules will be illustrated with the Nitrogen-vacancy center magnetometers. In the end, we will also 
introduce the numerical error evaluation and control tools in the package, which can be used to check and ensure 
the validity of the designed scheme. 

\section{Architecture overview}

\subsection{Package architecture and structure}

A typical scheme for quantum parameter estimation usually consists of four steps: state preparation, parameterization, 
quantum measurement, and classical parameter estimation~\cite{Braun2018,Pezze2018,Liu2020,Liu2022}. The mission of 
scheme design for quantum parameter estimation in a specific scenario is to provide the optimal forms of these 
four steps (or some of them if the scenario requires certain fixed steps). Hence, the core philosophy of developing 
QuanEstimation.jl is the efficient realization of scheme design for quantum parameter estimation in a given scenario. 
This philosophy indicates that the architecture of QuanEstimation.jl must be centered around the scheme, as shown 
in Fig.~\ref{fig:arch}(a). Therefore, defining a scheme is the first thing to do when using this package. Once the scheme 
is defined, the scheme evaluation, such as the calculations of various mathematical tools or metrological resources, and 
scheme design can be further performed. In the meantime, some utility tools like error evaluation/control and unit 
tests can also be executed. The error evaluation and control will be thoroughly discussed in Sec.~\ref{sec:accuracysec}. 
As a development tool, the unit tests would help the developers check the correctness and compatibility of the new codes. 
More details of it can be found in the documentation~\cite{doclink}. 

In QuanEstimation.jl, two methods can be used to define a scheme. The first and most common one is using the function 
\code{GeneralScheme()}, in which the basic elements of the scheme, such as the probe state, the parameterization process, 
and the measurement form, are manually input by the users. The system Hamiltonian, decay modes, and other systematic 
information are input when defining the parameterization process. The introduction and usage of this function will be 
thoroughly provided in Sec.~\ref{sec:schemeset}. Currently, the parameterization process can only be defined within 
the package. In the next version, we will include an interface to allow it, especially the dynamics process, to be 
defined via user-specific scripts that compact with other Julia ecosystems.

The second method to define a scheme is using the modules for specific quantum systems. These specific system modules 
are designed to improve the user experience and balance the efficiency and versatility of the package. In QuanEstimation.jl, 
each module is written as an independent package for the convenience of package development and management. The common 
parts used by all specific system modules are put into a base package named QuanEstimationBase.jl, as shown in Fig.~\ref{fig:arch}(b). 
All modules of specific systems can call the functions in this base package to further perform the scheme evaluation and design. 
Furthermore, the functions only available in this physical system and algorithms that are particularly efficient for it are 
also written inside the module to improve the computing efficiency. More details and demonstration of the modules will be 
given in Sec.~\ref{sec:module}. All modules and QuanEstimationBase are made up of the entire package of QuanEstimation.jl.

\subsection{Installation and calling}

As a registered Julia package, QuanEstimation.jl can be easily installed in Julia via the package Pkg.jl. The specific codes for its 
installation in the Read-Eval-Print Loop (REPL) are as follows: 
\begin{lstlisting}[label=lst:Scheme_object]
julia> using Pkg
julia> Pkg.add("QuanEstimation")
\end{lstlisting}
After the installation, it can be called in the REPL with the following codes: 
\begin{lstlisting}[label=lst:Scheme_object]
julia> using QuanEstimation
\end{lstlisting}
All the subpackages and functions can be directly applied once the codes in above line are used. In the demonstration codes the functions 
are called specifically (for example \code{using QuanEstimation:SigmaX}) to the purpose of reminding the readers which functions are belong 
to the package. 

\section{Scheme setup}
\label{sec:schemeset}

\subsection{General scheme}

A general scheme for quantum parameter estimation consists of four elements: probe state, parameterization process, 
measurement, and classical estimation strategy. These elements should be defined first when using QuanEstimation.jl. In the process 
of scheme design, these inputs, or some of them, work as the initial guesses of the optimizations. In QuanEstimation.jl, 
the scheme can be defined via the function \code{GeneralScheme()}, and all information in this function will be further 
used to construct a \code{struct} in Julia. The demonstration codes of its usage are as follows: 
\begin{lstlisting}[label=lst:Scheme_object]
julia> using QuanEstimation:GeneralScheme,QubitDephasing,PlusState,SIC
julia> dynamics = QubitDephasing([0.5,0.5,0.5],"z",0.1,0:0.01:1)
julia> GeneralScheme(;probe=PlusState(), param=dynamics, 
       measurement=SIC(2), x=nothing, p=nothing, dp=nothing)
\end{lstlisting}
In \code{GeneralScheme()}, the keyword arguments \code{probe=}, \code{param=}, and \code{measurement=} represent the input 
of the probe state, parameterization process, and measurement, respectively. In the case that a prior probability distribution exists, 
the regime of the arguments of the prior distribution, the distribution, and its derivatives should also be input via the keyword arguments 
\code{x}, \code{p}, and \code{dp}. The data types of the inputs are the same as those in the hybrid-language package, and more thorough 
demonstrations of them can be found in Ref.~\cite{Zhang2022}. In the following, we will briefly introduce how to define these elements in 
QuanEstimation.jl. 

In QuanEstimation.jl, the probe state is generally represented by a state vector or a density matrix. For a pure state, the 
data type of input probe state can either be a vector or a matrix and for a mixed state, it has to be a matrix. Several 
constantly used states are integrated into the package for convenience, as shown in Table~\ref{table:probestate}, and more 
will be involved in the future. The basis of these states is the same as the input Hamiltonian. If $(1,0)^{\mathrm{T}}:=\ket{0}$ 
and $(0,1)^{\mathrm{T}}:=\ket{1}$, then the functions \code{PlusState()} and \code{MinusState()} are in fact the states 
$(\ket{0}+\ket{1})/\sqrt{2}$ and $(\ket{0}-\ket{1})/\sqrt{2}$. \code{BellState(1)} to \code{BellState(4)} are the states 
$(\ket{00}+\ket{11})/\sqrt{2}$, $(\ket{00}-\ket{11})/\sqrt{2}$, $(\ket{01}+\ket{10})/\sqrt{2}$, and $(\ket{01}-\ket{10})/\sqrt{2}$, 
respectively. Some Julia packages like QuantumToolbox.jl~\cite{QTjl}, QuantumOptics.jl~\cite{Kramer2018}, and 
QuantumInformation.jl~\cite{Gawron2018} also contain many well-used quantum states, and the users can also call them for the 
generation of probe states. 

\begin{table}[tp]
\centering
\begin{tabular}{c|c}
\hline
\hline
Function name & Probe state \\
\hline
\rule{0pt}{3.5ex} PlusState() & $\left(\frac{1}{\sqrt{2}}, \frac{1}{\sqrt{2}}\right)^{\mathrm{T}}$ \\
\hline
\rule{0pt}{3.5ex} MinusState() & $\left(\frac{1}{\sqrt{2}}, -\frac{1}{\sqrt{2}}\right)^{\mathrm{T}}$ \\
\hline
\rule{0pt}{3.5ex} BellState(1) & $\left(\frac{1}{\sqrt{2}}, 0,0,\frac{1}{\sqrt{2}}\right)^{\mathrm{T}}$ \\
\hline
\rule{0pt}{3.5ex} BellState(2) & $\left(\frac{1}{\sqrt{2}}, 0,0,-\frac{1}{\sqrt{2}}\right)^{\mathrm{T}}$ \\
\hline
\rule{0pt}{3.5ex} BellState(3) & $\left( 0,\frac{1}{\sqrt{2}},\frac{1}{\sqrt{2}},0\right)^{\mathrm{T}}$ \\
\hline
\rule{0pt}{3.5ex} BellState(4) & $\left( 0,\frac{1}{\sqrt{2}},-\frac{1}{\sqrt{2}},0\right)^{\mathrm{T}}$ \\
\hline
\rule{0pt}{3.5ex} SigmaX() & $\left(\begin{array}{cc}
0 & 1 \\
1 & 0 \\
\end{array}\right)$ \\
\hline
\rule{0pt}{3.5ex} SigmaY() & $\left(\begin{array}{cc}
0 & -i \\
i & 0 \\
\end{array}\right)$ \\
\hline
\rule{0pt}{3.5ex} SigmaZ() & $\left(\begin{array}{cc}
1 & 0 \\
0 & -1 \\
\end{array}\right)$ \\
\hline
\hline
\end{tabular}
\caption{Some integrated quantum states and operators in QuanEstimation.jl.}
\label{table:probestate}
\end{table}

\begin{table*}[tp]
\centering
\begin{tabular}{c|c|c}
\hline
\hline
Functions & Arguments & Control shape \\
\hline
\rule{0pt}{3.5ex} ZeroCTRL() & \diagbox{}{} & $0$ \\
\hline
\rule{0pt}{3.5ex} LinearCTRL(k, c0) & k: $k$, c0: $c_0$ & $kt+c_0$\\
\hline
\rule{0pt}{3.5ex} SineCTRL(A, $\omega$, $\phi$) & A: $A$, $\omega$: $\omega$, $\phi$: $\phi$ & $A \sin(\omega t +\phi)$    \\
\hline
\rule{0pt}{3.5ex} SawCTRL(k, n) & k: $k$, n: $n$ & $2k\Big(\frac{nt}{T}-\lfloor 0.5+\frac{nt}{T}\rfloor\Big)$\\
\hline
\rule{0pt}{3.5ex} TriangleCTRL(k, n) & k: $k$, n: $n$ & $2\Big|2k\big(\frac{nt}{T}-\lfloor 0.5+\frac{nt}{T} \rfloor\big)\Big|-1 $\\
\hline
\rule{0pt}{3.5ex} GaussianCTRL(A, $\mu$, $\sigma$) & A: $A$, $\mu$: $\mu$, $\sigma$: $\sigma$ & $A e^{-(t-\mu)^2/(2\sigma)}$ \\
\hline
\rule{0pt}{3.5ex} GaussianEdgeCTRL(A, $\sigma$) & A: $A$, $\sigma$: $\sigma$ & $A-Ae^{-t^2/\sigma}-Ae^{-(t-T)^2/\sigma}$ \\
\hline
\hline
\end{tabular}
\caption{Currently available control functions in QuanEstimation.jl. In the expressions 
$T$ is the end time of the array \code{tspan}, and $\lfloor\cdot\rfloor$ denotes 
the floor function.} 
\label{table:ctrlamp}
\end{table*}

The parameterization process plays a critical role in quantum parameter estimation. In general, this process is realized 
by quantum dynamics. In QuanEstimation.jl, the focus is primarily on the dynamics governed by the master equation in the 
Lindblad form:
\begin{equation}
\partial_t\rho =-i[H,\rho]+\sum_i \gamma_i\left(\Gamma_i\rho\Gamma^{\dagger}_i
-\frac{1}{2}\left\{\rho,\Gamma^{\dagger}_i \Gamma_i \right\}\right),
\end{equation}
where $\rho$ is the evolved density matrix, $H$ is the total Hamiltonian, and $\Gamma_i$ and $\gamma_i$ are the $i$th decay 
operator and decay rate, respectively. $\gamma_i$ could either be constant (a float number) or time-dependent (a vector). 
In the meantime, QuanEstimation.jl can also deal with the non-dynamical processes, such as the quantum channels described 
by Kraus operators, i.e., $\rho=\sum_i K_i\rho_0 K_i^{\dagger}$, where $K_i$ is the $i$th Kraus operator satisfying 
$\sum_{i}K^{\dagger}_i K_i=\mathbb{I}$ with $\mathbb{I}$ the identity operator, and $\rho_0$ is the probe state.

In QuanEstimation.jl, the master equations can be defined via the function \code{Lindblad()}, and the quantum channels can be 
defined via the function \code{Kraus()}. The demonstration codes for calling these two functions are as follows: 
\begin{lstlisting}[label=lst:Scheme_parameterization]
julia> using QuanEstimation:Lindblad,SigmaX,SigmaY,SigmaZ,ZeroCTRL
julia> H0 = 0.5*SigmaX()+0.5*SigmaZ()
julia> dH = [SigmaZ()]
julia> tspan = 0:0.01:1
julia> Hc = [SigmaY()]
julia> decay = [[SigmaX(), 0.01]]
julia> dynamics = Lindblad(H0, dH, tspan, Hc, decay; ctrl=ZeroCTRL(), 
       dyn_method=:Ode)
\end{lstlisting}

\begin{lstlisting}[label=lst:Scheme_parameterization]
julia> using QuanEstimation:Kraus
julia> E0 = [1 0; 0 sqrt(0.5)]
julia> E1 = [sqrt(0.5) 0; 0 0]
julia> K = [E0, E1]
julia> dK = [[[0 0; 0 -0.5/sqrt(0.5)]], [[0 0.5/sqrt(0.5); 0 0]]]
julia> channel = Kraus(K, dK)
\end{lstlisting}

In the function \code{Lindblad()}, the argument \code{tspan} is an array representing the time length for the evolution. 
In general, the argument \code{H0} is a matrix or a vector of matrices representing the full Hamiltonian in the noncontrolled 
scheme or the free (noncontrolled) part of the Hamiltonian in the controlled scheme. It is a matrix when the Hamiltonian is 
time-independent and a vector with the length equivalent to that of \code{tspan} when it is time-dependent. The argument 
\code{dH} is a vector of matrices for time-independent Hamiltonians and a vector of vector of matrices for time-dependent 
Hamiltonians, which contains the derivatives of the Hamiltonian for the parameters to be estimated, same as that in the 
hybrid-language package~\cite{Zhang2022}. 

Moreover, the Hamiltonian and its derivative can also be defined by functions. This can be done with the help of the function 
\code{Hamiltonian()}, which takes the functions \code{H0(u)}, \code{dH(u)} and the values of \code{u} (a float number or a 
vector) as arguments. It is a multiparameter scenario when \code{u} is a vector. The output type of \code{H0(u)} should be a 
matrix, and that of \code{dH} should be a vector of matrices. In the case that the Hamiltonian is time-dependent, the functions 
should be in the form of \code{H0(u,t)} and \code{dH(u,t)}. The demonstration codes for calling 
\code{Lindblad()} with the functions \code{H0(u)} and \code{dH(u)} are as follows: 
\begin{lstlisting}[label=lst:Hamiltonian_function,mathescape]
julia> using QuanEstimation:SigmaX,SigmaZ,Hamiltonian,Lindblad
julia> H0(u) = (SigmaX()*cos(u)+SigmaZ()*sin(u))/2
julia> dH(u) = [(-SigmaX()*sin(u)+SigmaZ()*cos(u))/2]
julia> u = pi/4
julia> ham = Hamiltonian(H0, dH, u)
julia> decay = [[SigmaZ(), 0.01]]
julia> dynamics = Lindblad(ham, 0:0.01:1, decay)
\end{lstlisting}
The demonstration codes for the multiparameter scenario can be found in the documentation~\cite{doclink}. 

The argument \code{decay} is a vector containing the information of both decay operators and decay rates, and its input rule is 
\code{decay=[[Gamma1,gamma1], [Gamma2,gamma2],...]}, where \code{Gamma1} (\code{Gamma2}) and \code{gamma1} (\code{gamma2} ) 
represent the first (second) decay operator and decay rate, respectively. So do others, if there are any. Here the decay rate 
\code{gamma1} (\code{gamma2}) can either be a float number (representing a fixed decay rate) or a vector (representing a 
time-dependent decay rate), and when it is a vector, its length should be identical with \code{tspan}. The argument \code{Hc} 
is a vector of matrices representing the control Hamiltonians. Its default value is \code{nothing}, which represents the 
noncontrolled scheme. The argument \code{ctrl} is a vector of vectors containing the control amplitudes for the control 
Hamiltonians given in the argument \code{Hc}. Its default value \code{ZeroCTRL()} represents the zero control amplitudes for 
all control Hamiltonians. Some frequently used control amplitudes are integrated into the package for convenience, as given 
in Table~\ref{table:ctrlamp}, and more will be added in the future.  

In \code{Lindblad()}, the master equation is solved via the package DifferentialEquations.jl~\cite{Rackauckas2017} by default, 
and the corresponding setting is \code{dyn\_method=:Ode} (or \code{dyn\_method=:ode}). Alternatively, it can also be solved via the matrix 
exponential method by using \code{dyn\_method=:Expm} (or \code{dyn\_method=:expm}), which is suitable for small to medium-sized systems. 

In the function \code{Kraus()}, \code{K} and \code{dK} are vectors of matrices representing the Kraus operators and corresponding derivatives 
on the parameters to be estimated. Similar to the function \code{Lindblad()}, here the Kraus operators and their derivatives can also be defined 
as functions \code{K(u)} and \code{dK(u)}, of which the output types are also vector of matrices and vector of vector of matrices. In this case, 
the function \code{Kraus()} is called via 
the following format: 
\begin{lstlisting}[label=lst:Scheme_parameterization]
julia> using QuanEstimation:Kraus
julia> u = pi/4
julia> K(u) = [[1 0; 0 sqrt(1-u)], [0 sqrt(u); 0 0]]
julia> dK(u) = [[[0 0; 0 -0.5/sqrt(1-u)]], [[0 0.5/sqrt(u); 0 0]]]
julia> channel = Kraus(K, dK, u)
\end{lstlisting}
More demonstration codes of this case can be found in the documentation~\cite{doclink}. 

Regarding the measurement, the data type of the input measurement is a vector of matrices with each entry an element of a set of 
positive operator-valued measure (POVM). If no specific measurement is input, the rank-one symmetric informationally complete 
POVM (SIC-POVM) will be used as the default choice, which can also be manually invoked via the function \code{SIC()}.  

After the scheme setup is finished, the metrological quantities can be readily evaluated. All the metrological quantities 
given in the hybrid-language package~\cite{Zhang2022} are available in QuanEstimation.jl, such as the Quantum Cram\'{e}r-Rao 
bounds~\cite{Helstrom1976,Holevo1982} and various types of quantum Fisher information matrix (QFIM)~\cite{Liu2020}, Holevo 
Cram\'{e}r-Rao bound (HCRB)~\cite{Holevo1973,Rafal2020,Nagaoka1989,Hayashi2008}, and Nagaoka-Hayashi bound 
(NHB)~\cite{Nagaoka1989,Hayashi2005,Conlon2021}. Demonstration codes for their calculations in QuanEstimation.jl is as follows: 
\begin{lstlisting}[label=lst:metrological_quantities]
julia> using QuanEstimation:Lindblad,ZeroCTRL,GeneralScheme,PlusState
julia> using QuanEstimation:Hamiltonian,SIC,SigmaX,SigmaY,SigmaZ
julia> using QuanEstimation:QFIM,CFIM,HCRB,NHB
julia> using LinearAlgebra:I
julia> H0(u) = (SigmaX()*cos(u)+SigmaZ()*sin(u))/2
julia> dH(u) = [(-SigmaX()*sin(u)+SigmaZ()*cos(u))/2]
julia> ham = Hamiltonian(H0, dH, pi/4)
julia> dynamics = Lindblad(ham,0:0.01:1,[SigmaY()],[[SigmaZ(), 0.01]])
julia> scheme = GeneralScheme(; probe=PlusState(),param=dynamics,
       measurement=SIC(2))
julia> QFIM(scheme; LDtype=:SLD)
julia> CFIM(scheme)
julia> HCRB(scheme; W=I(1))
julia> NHB(scheme; W=I(1)) 
\end{lstlisting}
In the function \code{QFIM()}, the keyword argument \code{LDtype=:SLD} means that the calculated QFIM is based on the symmetric 
logarithmic derivative (SLD), same as that in the hybrid-language package. Details of calling other types of QFIM can be found 
in Ref.~\cite{Zhang2022}. In the functions \code{HCRB()} and \code{NHB()}, the argument \code{W} represents the weight matrix, 
and its definition can also be found in Ref.~\cite{Zhang2022}. In the case that a prior distribution exists, Bayesian types 
of QFIM or other tools like the Van Trees bound (VTB)~\cite{vanTrees1968}, and its quantum version (QVTB), also known as 
Tsang-Wiseman-Caves bound~\cite{Tsang2011}, should be used for the evaluation of precision limit. Here we present the 
demonstration codes for the calculations of VTB and QVTB: 
\begin{lstlisting}[label=lst:metrological_quantities_bayesian]
julia> using QuanEstimation:Lindblad,GeneralScheme,Hamiltonian
julia> using QuanEstimation:SigmaX,SigmaZ,PlusState,SIC,VTB,QVTB
julia> xspan = range(0, pi; length=200)
julia> mu = pi/2
julia> p = xspan .|> xspan->exp(-(xspan-mu)^2/2)
julia> dp = xspan .|> xspan->-(xspan-mu)*exp(-(xspan-mu)^2/2)
julia> H0(x) = (SigmaX()*cos(x)+SigmaZ()*sin(x))/2
julia> dH(x) = [(-SigmaX()*sin(x)+SigmaZ()*cos(x))/2]
julia> ham = Hamiltonian(H0, dH, pi/4)
julia> dynamics = Lindblad(ham, 0:0.01:1, [[SigmaZ(), 0.01]])
julia> scheme = GeneralScheme(; probe=PlusState(), param=dynamics, 
       measurement=SIC(2), x=xspan, p=p, dp=dp)
julia> VTB(scheme)
julia> QVTB(scheme)
\end{lstlisting}
The outputs of all these functions are vectors representing the time evolutions of the metrological quantities. 

The advantage of using \code{GeneralScheme()} is the flexible choice of physical systems. In principle, any quantum system 
can be implemented in QuanEstimation.jl to evaluate the metrological quantities or perform scheme design, regardless of the 
computing efficiency. However, many physicists mainly focus on certain specific physical systems. The convenience of scheme 
setup and computing efficiency of scheme design for these systems are critical to them, which may not be fully satisfying 
with the function \code{GeneralScheme()}. For the sake of improving the efficiency of scheme setup and scheme design for 
certain specific systems, the modules are developed in QuanEstimation.jl, which is a significant feature that did not appear 
in the last version of the hybrid-language package~\cite{Zhang2022}, and will be thoroughly introduced in the next section. 

\subsection{Modules for specific systems}
\label{sec:module}

Many quantum systems have present significant advantages in various scenarios of quantum parameter estimation, such as the 
Nitrogen-vacancy centers~\cite{Degen2017}, cold atoms~\cite{Huang2014,Pezze2018}, trapped ions~\cite{Marciniak2022}, 
optomechanical systems~\cite{Zhu2023,Wu2023}, and quantum circuits~\cite{,Marciniak2022,Su2017,Kaubruegger2023}. 
For these specific quantum systems, the Hamiltonian structures and the parameters to be estimated are usually fixed. Hence, 
for the researchers focusing on a specific quantum system, especially those not experienced in Julia or even coding, a module 
that integrates the Hamiltonian and other information of this system would make QuanEstimation.jl more user-friendly. In the 
meantime, a module for a specific physical system would also make it easier to adjust the codes according to the features of 
this system and implement the algorithms that are particularly efficient for the scheme design in this system. The computing 
efficiency of the scheme design would then be improved. 

All specific system modules will be structurally written as independent packages for the convenience of development and maintenance. 
However, they can be directly called when \code{using QuanEstimation} is applied. Currently, most modules are still under construction 
and will be available in both the hybrid-language package and Julia package in a short time. In this paper we take the Nitrogen-vacancy 
center magnetometer as an example of the modules and demonstrate its usage. Some other packages like qsensoropt~\cite{Belliardo2024,
Belliardo2024a} can also be used to design magnetometer with Nitrogen-vacancy center. The Hamiltonian of the Nitrogen-vacancy center 
is~\cite{Barry2020,Felton2009,Schwartz2018,Rembold2020}
\begin{equation}
H_0/\hbar=DS^2_3+g_{\mathrm{S}}\vec{B}\cdot\vec{S}+g_{\mathrm{I}}\vec{B}\cdot\vec{I}
+\vec{S}^{\,\mathrm{T}}\mathcal{A}\vec{I},
\label{eq:NV_H}
\end{equation}
where $S_i=s_i\otimes\mathbb{I}$ and $I_i=\mathbb{I}\otimes\sigma_i$ ($i=1,2,3$) are the electron and nuclear ($^{15}\mathrm{N}$) 
operators. $s_1$, $s_2$ and $s_3$ are spin-1 operators with the expressions 
\begin{eqnarray}
s_1 &=& \frac{1}{\sqrt{2}}\left(\begin{array}{ccc}
0 & 1 & 0 \\
1 & 0 & 1 \\
0 & 1 & 0
\end{array}\right), \\
s_2 &=& \frac{1}{\sqrt{2}}\left(\begin{array}{ccc}
0 & -i & 0\\
i & 0 & -i\\
0 & i & 0
\end{array}\right), \\
s_3 &=& \left(\begin{array}{ccc}
1 & 0 & 0\\
0 & 0 & 0\\
0 & 0 & -1
\end{array}\right).
\end{eqnarray}
The vectors $\vec{S}=(S_1,S_2,S_3)^{\mathrm{T}}$, $\vec{I}=(I_1,I_2,I_3)^{\mathrm{T}}$, and $\mathcal{A}$ is the hyperfine tensor, and 
in this case $\mathcal{A}=\mathrm{diag} (A_1,A_1,A_2)$ with $A_1$ and $A_2$ the axial and transverse magnetic hyperfine coupling 
coefficients. The hyperfine coupling between the magnetic field and the electron is approximated to be isotopic. 
$g_{\mathrm{S}}=g_\mathrm{e}\mu_\mathrm{B}/\hbar$ and $g_{\mathrm{I}}=g_\mathrm{n}\mu_\mathrm{n}/\hbar$ with 
$g_\mathrm{e}$ ($g_\mathrm{n}$) the $g$ factor of the electron (nuclear), $\mu_\mathrm{B}$ ($\mu_\mathrm{n}$) 
the Bohr (nuclear) magneton, and $\hbar$ the Plank's constant. $\vec{B}$ is the external magnetic field 
that needs to be estimated. In this system, the control Hamiltonian can be expressed by 
\begin{equation}
H_{\mathrm{c}}/\hbar=\sum^3_{i=1}\Omega_i(t)S_i, 
\end{equation} 
where $\Omega_i(t)$ is a time-varying Rabi frequency. Due to the fact that the electron suffers from the noise of
dephasing in practice, the dynamics of the Nitrogen-vacancy center is then described by 
\begin{equation}
\partial_t\rho=-i[H_0+H_{\mathrm{c}},\rho]+\frac{\gamma}{2}(S_3\rho S_3-S^2_3\rho-\rho S^2_3),
\end{equation}
where $\gamma$ is the dephasing rate, which is usually inversely proportional to the dephasing time $T^{*}_2$. 

\begin{figure*}[t]
\centering\includegraphics[width=15.cm]{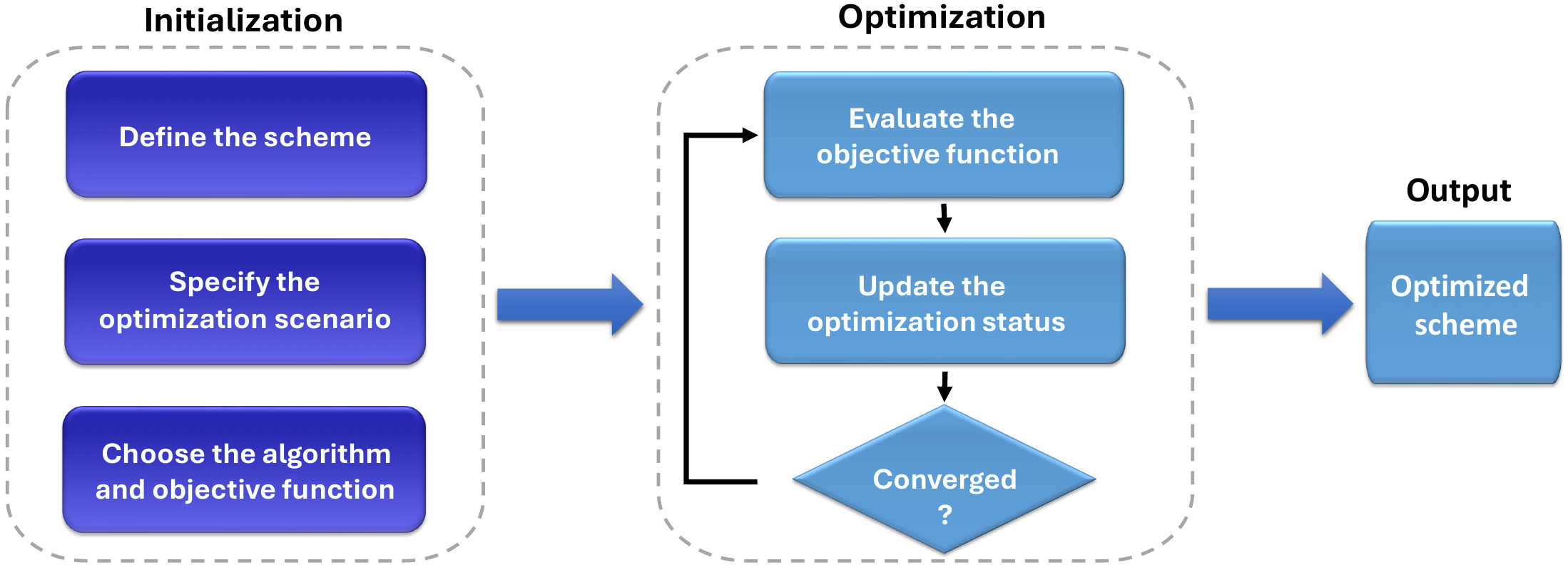}
\caption{The process of scheme design in QuanEstimation.jl, which includes three steps: 
(1) initialization, (2) optimization, and (3) output. 
\label{fig:opt_scheme}}
\end{figure*}

In QuanEstimation.jl, this module can be used to define the scheme by calling the function \code{NVMagnetometerScheme()}. The 
coefficients of the Hamiltonian in the module are taken as those given in Refs.~\cite{Barry2020,Felton2009}. After the scheme 
is defined, the scheme evaluation and design can be further applied. The demonstration codes for calling this module and 
evaluating the value of QFI are as follows: 
\begin{lstlisting}[label=lst:Scheme,mathescape]
julia> using QuanEstimation:NVMagnetometerScheme,QFIM
julia> scheme = NVMagnetometerScheme()
NVMagnetometerScheme
├─ StatePreparation => DensityMatrix
│  ├─ ndim = (6,)
│  └─ $\psi$0  = [0.7071067811865475, 0.0, 0.0, 
│             0.0, 0.7071067811865475, 0.0]
├─ Parameterization => LindbladDynamics
│  ├─ tspan = 0.0:0.01:2.0
│  ├─ Hamiltonian => NVCenterHamiltonian
│  │  ├─ D  = 18032.741831605414
│  │  ├─ gS = 176.1176841602438
│  │  ├─ gI = 0.027143360527015815
│  │  ├─ A1 = 22.933626371205488
│  │  ├─ A2 = 19.038051480754145
│  │  └─ B  = [0.5, 0.5, 0.5]
│  ├─ Controls
│  │  ├─ Hc = [S1, S2, S3]
│  │  └─ ctrl = nothing
│  └─ decays
│     ├─ decay_opt => [S3]
│     └─ $\gamma$ = 6.283185307179586
└─ Measurement
    └─ M = nothing
julia> F = QFIM(scheme)
\end{lstlisting}

The progress of other modules, such as the Mach-Zehnder interferometer, cold atoms, trapped ions, optomechanical systems, 
and quantum circuits, will be constantly updated in the documentation~\cite{doclink}. 

\section{Scheme design}

Scheme design is the major mission of QuanEstimation.jl, and also the key reason why Julia is used to write the computational 
core of the hybrid-language package~\cite{Zhang2022}. In QuanEstimation.jl, the process of scheme design consists of three 
steps, including initialization, optimization, and output, as shown in Fig.~\ref{fig:opt_scheme}. Three elements are required 
in the first step. Apart from defining the scheme, the optimization scenario must be specified, and the objective function 
and optimization algorithm should be chosen. 

As discussed in the previous section, the scheme can be defined via the general approach or specific system modules. Specifying 
the optimization scenario means the user needs to clarify which part of the scheme needs to be optimized. Currently, the package 
includes probe state optimization, control optimization, measurement optimization, and comprehensive optimization, same as those 
in the hybrid-language package. Detailed introduction of them can be found in Ref.~\cite{Zhang2022}. The probe state, control, 
measurement, and comprehensive optimization can be specified via the functions \code{StateOpt()}, \code{ControlOpt()}, 
\code{MeasurementOpt()}, and \code{CompOpt()}. In the case of measurement optimization, QuanEstimation.jl now provides three 
types of scenarios, including rank-one projective measurements, linear combinations, and rotations of a set of input measurements. 
The desired optimization strategy can be selected by setting \code{mtype=:Projection}, \code{mtype=:LC}, or 
\code{mtype=:Rotation}, respectively. Regarding the comprehensive optimization, four types of joint optimization, including 
the probe state and measurement (realized by setting \code{type=:SM}), probe state and control (\code{type=:SC}), control and 
measurement (\code{type=:CM}), and all three variables together (\code{type=:SCM}), can be executed. 

After specifying the optimization scenario, the objective function and algorithm for optimization should be chosen. This part 
can be neglected if the user has no preference on the objective function and algorithm since all scenarios have default choices. 
Most metrological tools can be taken as the objective function. QuanEstimation.jl includes both gradient-based algorithms, such 
as the gradient ascent pulse engineering algorithm and its advanced version based on automatic differentiation, and gradient-free 
algorithms such as particle swarm optimization and differential evolution. Details of the available algorithms and objective 
functions in each scenario can be found in Ref.~\cite{Zhang2022}. 

The information created in the step of initialization is stored as a \code{struct} in QuanEstimation.jl, which not only contains 
the information of the scheme but also the supplementary information that can assist the precision analysis of the optimized 
scheme, such as the number of iterations used during the optimization process and the convergence criteria of optimization.

Once the initialization is finished, the scheme is then ready to be optimized. In the step of optimization, the scheme data are 
updated by the selected optimization algorithm, and the objective function is evaluated, as shown in Fig.~\ref{fig:opt_scheme}. 
This process continues until the convergence conditions are met. After the value of the objective function is converged, the 
optimized scheme is then output. The data will be saved into files (HDF5 format with extension name .dat) via the JLD2 package 
and printed on the screen. 

Here we provide the demonstration codes for the scenario of control optimization: 
\begin{lstlisting}[label=lst:Copt]
julia> using QuanEstimation:NVMagnetometerScheme
julia> using QuanEstimation:ControlOpt,optimize!,autoGRAPE
julia> scheme = NVMagnetometerScheme()
julia> opt = ControlOpt()
julia> optimize!(scheme, opt; algorithm=autoGRAPE(), savefile=true)
\end{lstlisting}
More examples and demonstrations of other scenarios can be found in the documentation~\cite{doclink}. 

Adaptive measurement is another well-used scenario in quantum parameter estimation~\cite{Berry2000,Berry2001,Hentschel2010,
Hentschel2011,Lovett2013,Rambhatla2020,Berni2015,Liu2022}, in which a vector of tunable parameters is used to enhance the 
measurement precision. Similar to the hybrid-language package, QuanEstimation.jl can also realize the adaptive measurements. 
To do it, the function \code{AdaptiveStrategy()} should be used first before defining the scheme to claim the regime of the 
parameters to be estimated, the prior distribution, and its derivatives to the parameters. The process of scheme setup is 
the same as other scenarios, which can be realized via \code{GeneralScheme()} or certain specific system modules. Then the 
function \code{adapt!()} is called to perform the adaptive measurement. The demonstration codes for the adaptive 
measurement are as follows: 
\begin{lstlisting}[label=lst:Adapt]
julia> using QuanEstimation:GeneralScheme,Hamiltonian,Lindblad
julia> using QuanEstimation:PlusState,SIC,SigmaX,SigmaZ
julia> using QuanEstimation:AdaptiveStrategy,adapt!
julia> xspan = range(0, pi; length=200)
julia> mu = pi/2
julia> p = xspan .|> xspan->exp(-(xspan-mu)^2/2)
julia> dp = xspan .|> xspan->-(xspan-mu)*exp(-(xspan-mu)^2/2)
julia> strat = AdaptiveStrategy(; x=xspan, p=p, dp=dp)
julia> H0(x) = (SigmaX()*cos(x)+SigmaZ()*sin(x))/2
julia> dH(x) = [(-SigmaX()*sin(x)+SigmaZ()*cos(x))/2]
julia> ham = Hamiltonian(H0, dH, pi/4)
julia> dynamics = Lindblad(ham, 0:0.01:1, [[SigmaZ(), 0.01]])
julia> scheme = GeneralScheme(; probe=PlusState(), param=dynamics, 
       measurement=SIC(2), strat=strat)
julia> adapt!(scheme; method="FOP", savefile=false, max_episode=1000)
\end{lstlisting}
More details on the usage of adaptive measurement are given in Ref.~\cite{Zhang2022} and the documentation~\cite{doclink}. 
The online and offline adaptive phase estimations in the Mach-Zehnder interferometer are integrated into the module of 
Mach-Zehnder interferometer, and will be thoroughly introduced in another paper. 

\begin{figure*}[tp]
\centering\includegraphics[width=15.cm]{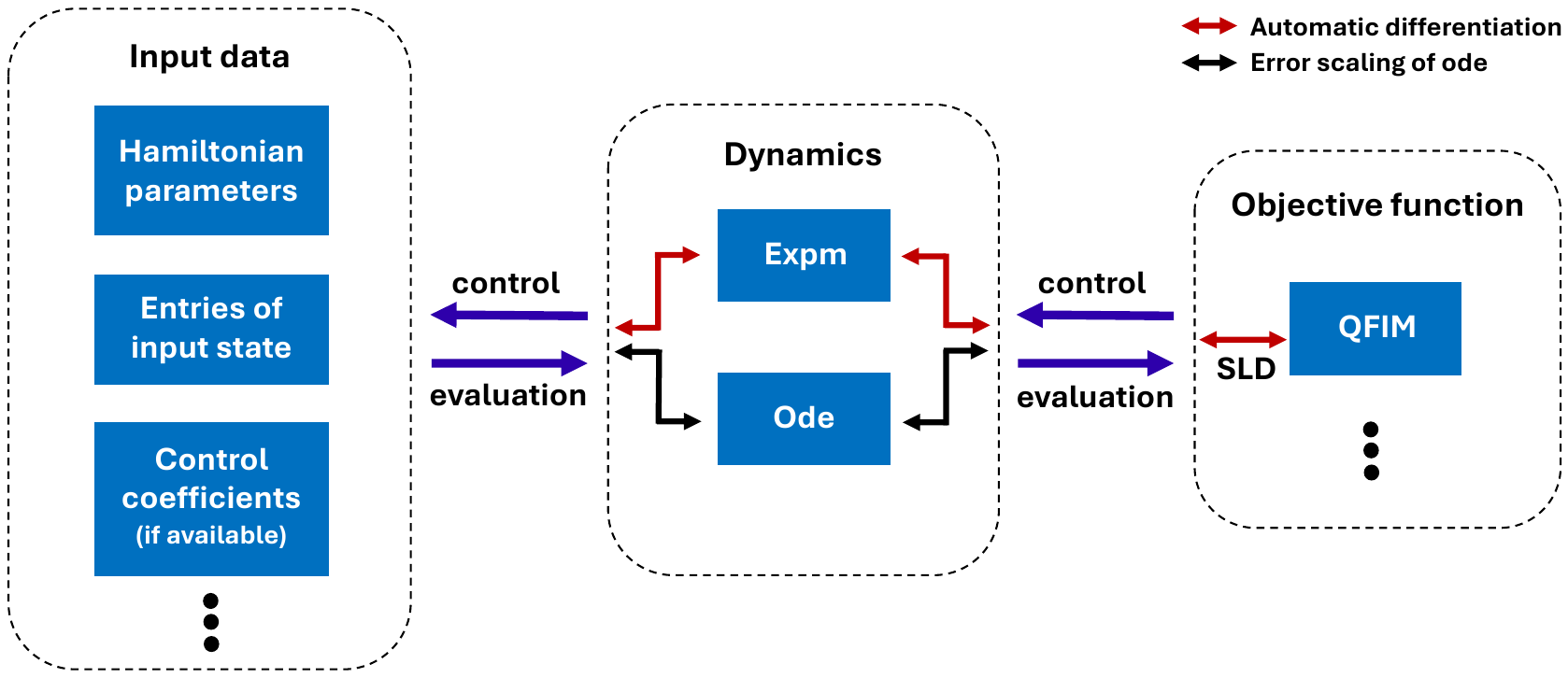}
\caption{Processes of computational error evaluation and control in QuanEstimation.jl. 
\label{fig:error}}
\end{figure*}

\section{Numerical error evaluation and control}
\label{sec:accuracysec}

To evaluate the numerical precision of the output scheme or the calculated mathematical tools, the computational errors 
should be evaluated, and sometimes even controlled. QuanEstimation.jl provides two functions \code{error\_evaluation()} 
and \code{error\_control()} to help the users perform the error evaluation and control. The error evaluation uses the 
given precision of the input data to evaluate the error scaling of the output, and the error control uses the given error 
of the output to provide a suggested precision scaling of the input data. The demonstration codes for their usage are 
given as follows:
\begin{lstlisting}[label=lst:error_evaluation]
julia> using QuanEstimation:NVMagnetometerScheme
julia> using QuanEstimation:error_evaluation,error_control
julia> scheme = NVMagnetometerScheme()
julia> error_evaluation(scheme; input_error_scaling=1e-8, 
	  objective="QFIM", SLD_eps=1e-8)
julia> error_control(scheme; output_error_scaling=1e-6, 
	  objective="QFIM", SLD_eps=1e-8)
\end{lstlisting}

In QuanEstimation.jl, the total error of the output, such as the QFIM, is evaluated via the error propagation relation 
\begin{equation}
\delta f=\sqrt{\sum_{i}\left(\frac{\partial f}{\partial x_i} \right)^2\delta^2 x_i},    
\label{eq:error_propa}
\end{equation}
where $f=f(x_1,x_2,\cdots)$ is the output with $x_i$ the $i$th input parameter. $\delta f$ and $\delta x_i$ represent 
the errors of $f$ and $x_i$, respectively. The specific processes of error evaluation and control in the package are 
illustrated in Fig.~\ref{fig:error}. 

The function \code{error\_evaluation()} provides the evaluated error scaling of the output at the final evolution time 
according to the given precision of the input data. In the evaluation process, the precision of the input data is assumed 
to be the same, which can be set via the key argument \code{input\_error\_scaling} in the function. In practice, if not 
all the input data can be set to the same precision, the input precision used for error evaluation can be taken as the 
worst one among them. In the case that the dynamics is calculated by the method \code{Expm} (\code{dyn\_method=:Expm}), 
the gradients in Eq.~(\ref{eq:error_propa}) are all evaluated via automatic differentiation through chain rules, as shown 
in Fig.~\ref{fig:error}. When the dynamics is solved by \code{Ode} (\code{dyn\_method=:Ode}), the errors of the evolved 
state and its derivatives are roughly evaluated as the summation of the input precision scaling and the time step to the 
fourth power, which is due to the fact that the global error of Tsitouras 5/4 Runge-Kutta method can be roughly expressed 
by $\mathcal{O}(h^4)$~\cite{Tsitouras2011} with $h$ the time step. In the case of adaptive timestepping, the largest time 
step is used to evaluate the global error. Next, the evolved state and its derivatives are taken as the new input and the 
error of output is further evaluated via the error propagation relation, in which the gradients are also evaluated by 
automatic differentiation. 

In the case that the output is the QFIM, the machine epsilon (set by the keyword argument \code{SLD\_eps}) in the 
calculation of SLD would also contribute to the final error. Here the machine epsilon means that if an eigenvalue of 
the density matrix is less than the given value, it will be truncated to zero in the calculation of SLD~\cite{Zhang2022}. 
A proper setting of \code{SLD\_eps} would help to improve the calculation stability of the QFIM. When the function 
\code{error\_evaluation()} is executed, the difference between the QFIMs before and after the truncation is applied will 
be shown on the screen (denoted by $\delta F$). If this value is too large, the value of \code{SLD\_eps} should be reset. 

The function \code{error\_control()} provides a suggested precision scaling of the input data based on the required error 
scaling of the output, which is set by the argument \code{output\_error\_scaling}. We still assume the precision of all 
input data is the same in this function. In practice, the users can take this suggested precision as the worst precision 
requirement for all input data. As long as the precision of all inputs is higher than the suggested scaling, the error of 
output would meet the user's requirement. In the case that \code{Expm} is applied, the suggested input precision scaling 
is fully evaluated via Eq.~(\ref{eq:error_propa}). When \code{Ode} is applied, the required error of the evolved state is 
also evaluated by automatic differentiation, and then the suggested input precision is calculated as the difference between 
the error of the evolved state and the (largest) time step to the fourth power. 

\section{Future developments and extensions}

Alongside enhancing the general computational efficiency of QuanEstimation.jl, our development roadmap prioritizes the 
development of modules for specific quantum systems or scenarios, including but not limited to the abstract models like the 
SU(2) and SU(1,1) interferometers, various toy models in quantum optics, quantum networks, and specific physical systems 
like quantum circuits, trapped ions, and cold atoms. Each module would be constructed as an independent subpackage. 
Concurrently, integration of more cutting-edge methodologies in quantum information and condensed-matter physics, and even 
those in computing science is also an important content for the development in the future. 

Recent advancements in quantum experimental technologies have enabled the automation of numerous experimental processes. 
In the meantime, the strategic vision driving the development of QuanEstimation and QuanEstimation.jl centers on the automatic 
realization of scheme design in quantum parameter estimation. Building upon this vision, another key development direction is 
the seamless integration of intelligent scheme design with experimental execution. This adaptive framework will enable dynamic 
scheme optimization through real-time environmental feedback, with redesigned configurations being autonomously deployed to 
maintain optimal estimation performance under varying conditions.

\section{Summary}

The development of QuanEstimation.jl aims at efficient scheme evaluation and design for quantum parameter estimation. 
This package can work as the computational core of its hybrid-language counterpart or as an independent package. 
The usage of QuanEstimation.jl is based on the construction of schemes. Once a scheme is constructed, all the metrological 
quantities discussed in Ref.~\cite{Zhang2022} can be evaluated and the optimal schemes can be provided according to the 
user's requirements. To balance the versatility and efficiency, we introduce modules in the package for specific physical 
systems and demonstrate its usage with the Nitrogen-vacancy center magnetometer. The package version of QuanEstimation.jl 
with respect to the contents of this paper is v0.2. The source codes can be found in GitHub~\cite{gitlink} and the documentation 
is available in the link in Ref.~\cite{doclink}. 

\section{Declaration of competing interest}
The authors declare that they have no conflicts of interest in this work.

\section{Acknowledgments}
The authors would like to thank all anonymous reviewers for their insightful suggestions on both the package and the presentation 
of this paper. Moreover, the authors would like to thank Dr.~Mao Zhang for her significant contributions to the coding, and would 
also like to thank Mr.~Zheng-Wei An, and Mr.~Xin-Ze Yan for their help on the coding and useful suggestions. This work was 
supported by the National Natural Science Foundation of China through Grant No.\,12175075.


\begin{thebibliography}{00}

\bibitem{Beloy2021}
Boulder Atomic Clock Optical Network (BACON) Collaboration, 
Frequency ratio measurements at 18-digit accuracy using an optical clock network, 
\href{https://doi.org/10.1038/s41586-021-03253-4}
{Nature \textbf{591} (7851) (2021) 564–569.}

\bibitem{Bao2020}
H. Bao, J. Duan, S. Jin, et al.,
Spin squeezing of 1011 atoms by prediction and retrodiction measurements,  
\href{https://doi.org/10.1038/s41586-020-2243-7}
{Nature \textbf{581} (7807) (2020) 159–163.}

\bibitem{Helstrom1976}
C. W. Helstrom, \emph{Quantum Detection and Estimation Theory} (Academic, New York, 1976).

\bibitem{Holevo1982}
A. S. Holevo, \emph{Probabilistic and Statistical Aspects of Quantum Theory} (North-Holland, Amsterdam, 1982).

\bibitem{Zhang2022}
M. Zhang, H.-M. Yu, H. Yuan, et al.,
QuanEstimation: An open-source toolkit for quantum parameter estimation, 
\href{https://doi.org/10.1103/PhysRevResearch.4.043057}
{Phys. Rev. Res. \textbf{4} (4) (2022) 043057.}

\bibitem{Bezanson2012}
J. Bezanson, S. Karpinski, V. B. Shah, et al., 
Julia: A fast dynamic language for technical computing,
\href{https://doi.org/10.48550/arXiv.1209.5145}
{arXiv:1209.5145}.

\bibitem{QTjl}
\href{https://github.com/qutip/QuantumToolbox.jl}
{\small{https://github.com/qutip/QuantumToolbox.jl}}

\bibitem{Kramer2018}
 S. Kr{\"a}mer, D. Plankensteiner, L. Ostermann, et al.,
QuantumOptics.jl: A Julia framework for simulating open quantum systems,
\href{https://doi.org/10.1016/j.cpc.2018.02.004}
{Comput. Phys. Commun. \textbf{227} (2018) 109-116.}

\bibitem{Gawron2018}
P. Gawron, D. Kurzyk, and {\L}. Pawela, 
QuantumInformation.jl-A Julia package for numerical computation in quantum information theory, 
\href{https://doi.org/10.1371/journal.pone.0209358}
{PLoS ONE \textbf{13} (12) (2018) e0209358.}

\bibitem{Yaojl}
X.-Z. Luo, J.-G. Liu, P. Zhang, et al.,  
Yao.jl: Extensible, Efficient Framework for Quantum Algorithm Design, 
\href{https://doi.org/10.22331/q-2020-10-11-341}
{Quantum \textbf{4} (2020) 341.}

\bibitem{QCjl}
\href{https://github.com/JuliaQuantumControl/QuantumControl.jl}
{\small{https://github.com/JuliaQuantumControl/QuantumControl.jl}}

\bibitem{Braun2018}
D. Braun, G. Adesso, F. Benatti, et al.,
Quantum-enhanced measurements without entanglement, 
\href{https://doi.org/10.1103/RevModPhys.90.035006}
{Rev. Mod. Phys. \textbf{90} (3) (2018) 035006.}

\bibitem{Pezze2018}
L. Pezz{\`e}, A. Smerzi, M. K. Oberthaler, et al., 
Quantum metrology with nonclassical states of atomic ensembles,
\href{https://doi.org/10.1103/RevModPhys.90.035005}
{Rev. Mod. Phys. \textbf{90} (3) (2018) 035005.}

\bibitem{Liu2020}
J. Liu, H. Yuan, X.-M. Lu, et al., 
Quantum Fisher information matrix and multiparameter estimation,
\href{https://doi.org/10.1088/1751-8121/ab5d4d}
{J. Phys. A: Math. Theor. \textbf{53} (2) (2020) 023001.}

\bibitem{Liu2022}
J. Liu, M. Zhang, H. Chen, et al., 
Optimal Scheme for Quantum Metrology, 
\href{https://doi.org/10.1002/qute.202100080}
{Adv. Quantum Technol. \textbf{5} (2022) 2100080.}

\bibitem{Rackauckas2017}
C. Rackauckas and Q. Nie,
DifferentialEquations.jl – A Performant and Feature-Rich Ecosystem for Solving Differential Equations in Julia,
\href{https://doi.org/10.5334/jors.151}
{J. Open Res. Software \textbf{5} (1) (2017) 15.}

\bibitem{Holevo1973}
A. S. Holevo,
Statistical decision theory for quantum systems,
\href{https://doi.org/10.1016/0047-259X(73)90028-6}
{J. Multivariate Anal. \textbf{3} (4) (1973) 337-394.}

\bibitem{Rafal2020}
R. Demkowicz-Dobrza\'{n}ski, W. G\'{o}recki, and M. Gu\c{t}\u{a},
Multi-parameter estimation beyond Quantum Fisher Information,
\href{https://doi.org/10.1088/1751-8121/ab8ef3}
{J. Phys. A: Math. Theor. \textbf{53} (36) (2020) 363001.}

\bibitem{Hayashi2008}
M. Hayashi and K. Matsumoto,
Asymptotic performance of optimal state estimation in qubit system,
\href{https://doi.org/10.1063/1.2988130}
{J. Math. Phys. \textbf{49} (10) (2008) 102101.}

\bibitem{Nagaoka1989}
H. Nagaoka,
\emph{A new approach to Cramér-Rao bounds for quantum state estimation, in Asymptotic Theory Of
Quantum Statistical Inference: Selected Papers}
(World Scientific, Singapore, 2005), pp. 100-112.

\bibitem{Hayashi2005}
M. Hayashi, editor, \emph{Asymptotic Theory of Quantum Statistical Inference: Selected Papers}
(World Scientific, Singapore, 2005).

\bibitem{Conlon2021}
L. O. Conlon, J. Suzuki, P. K. Lam, et al., 
Efficient computation of the Nagaoka-Hayashi bound for multiparameter
estimation with separable measurements,
\href{https://doi.org/10.1038/s41534-021-00414-1}
{npj Quantum Inf. \textbf{7} (2021) 110.}

\bibitem{vanTrees1968}
H. L. Van Trees, \emph{Detection, estimation, and modulation theory: Part I}
(Wiley, New York, 1968).

\bibitem{Tsang2011}
M. Tsang, H. M. Wiseman, and C. M. Caves,
Fundamental Quantum Limit to Waveform Estimation,
\href{https://doi.org/10.1103/PhysRevLett.106.090401}
{Phys. Rev. Lett. \textbf{106} (9) (2011) 090401.}

\bibitem{Degen2017}
C. L. Degen, F. Reinhard, and P. Cappellaro, 
Quantum sensing,
\href{https://doi.org/10.1103/RevModPhys.89.035002}
{Rev. Mod. Phys. \textbf{89} (3) (2017) 035002.}

\bibitem{Huang2014}
J. Huang, M. Zhuang, and C. Lee,
Entanglement-enhanced quantum metrology: from standard quantum limit to Heisenberg limit,
\href{https://doi.org/10.1063/5.0204102}
{Appl. Phys. Rev. 11 (3) (2024) 031302. }

\bibitem{Marciniak2022}
C. D. Marciniak, T. Feldker, I. Pogorelov, et al., 
Optimal metrology with programmable quantum sensors,
\href{https://doi.org/10.1038/s41586-022-04435-4}
{Nature \textbf{603} (7902) (2022) 604-609.}

\bibitem{Zhu2023}
G.-L. Zhu, C.-S. Hu, Y. Wu, et al., Cavity optomechanical chaos, 
\href{https://doi.org/10.1016/j.fmre.2022.07.012}
{Fundam. Res. \textbf{3} (1) (2023) 63–74.}

\bibitem{Wu2023}
M. Wu, T. Tian, Z. Wang, 
Vibration induced transparency: Simulating an optomechanical system via the cavity QED setup with a movable atom, 
\href{https://doi.org/10.1016/j.fmre.2022.09.00}
{Fundam. Res. \textbf{3} (1) (2023) 50–56.}

\bibitem{Su2017}
Z.-E. Su, Y. Li, P. P. Rohde, et al., 
Multiphoton Interference in Quantum Fourier Transform Circuits and Applications to Quantum Metrology,
\href{https://doi.org/10.1103/PhysRevLett.119.080502}
{Phys. Rev. Lett. \textbf{119} (8) (2017) 080502.}

\bibitem{Kaubruegger2023}
R. Kaubruegger, A. Shankar, D. V. Vasilyev, et al.,
Optimal and Variational Multiparameter Quantum Metrology and Vector-Field Sensing, 
\href{https://doi.org/10.1103/PRXQuantum.4.020333}
{PRX Quantum \textbf{4} (2) (2023) 020333.}

\bibitem{Belliardo2024}
F. Belliardo, F. Zoratti, F. Marquardt, et al., 
Model-aware reinforcement learning for high-performance Bayesian experimental design in quantum metrology, 
\href{https://doi.org/10.22331/q-2024-12-10-1555}
{Quantum \textbf{8} (2024) 1555.}

\bibitem{Belliardo2024a}
F. Belliardo, F. Zoratti, and V. Giovannetti, 
Applications of model-aware reinforcement learning in Bayesian quantum metrology, 
\href{https://doi.org/10.1103/PhysRevA.109.062609}
{Phys. Rev. A \textbf{109} (6) (2024) 062609.}

\bibitem{Barry2020}
J. F. Barry, J. M. Schloss, E. Bauch, et al., 
Sensitivity optimization for NV-diamond magnetometry,
\href{https://doi.org/10.1103/RevModPhys.92.015004}
{Rev. Mod. Phys. \textbf{92} (1) (2020) 015004.}

\bibitem{Felton2009}
S. Felton, B. L. Cann, A. M. Edmonds, et al., 
Electron paramagnetic resonance studies of nitrogen interstitial defects in diamond, 
\href{https://doi.org/10.1088/0953-8984/21/36/364212}
{J. Phys.: Condens. Matter \textbf{21} (36) (2009) 364212.}

\bibitem{Schwartz2018}
I. Schwartz, J. Scheuer, B. Tratzmiller, et al., 
Robust optical polarization of nuclear spin baths using Hamiltonian engineering of nitrogen-vacancy
center quantum dynamics,
\href{https://doi.org/10.1126/sciadv.aat8978}
{Sci. Adv. \textbf{4} (2018) eaat8978.}

\bibitem{Rembold2020}
P. Rembold, N. Oshnik,  M. M. M\"{u}ller, et al., 
Introduction to quantum optimal control for quantum sensing with nitrogen-vacancy centers in diamond,
\href{https://doi.org/10.1116/5.0006785}
{AVS Quantum Sci. \textbf{2} (2020) 024701.}

\bibitem{Berry2000}
D. W. Berry and H. M. Wiseman, 
Optimal States and Almost Optimal Adaptive Measurements for Quantum Interferometry, 
\href{https://doi.org/10.1103/PhysRevLett.85.5098}
{Phys. Rev. Lett. \textbf{85} (24) (2000) 5098.}

\bibitem{Berry2001}
D. W. Berry, H. M. Wiseman, and J. K. Breslin, 
Optimal input states and feedback for interferometric phase estimation, 
\href{https://doi.org/10.1103/PhysRevA.63.053804}
{Phys. Rev. A \textbf{63} (5) (2001) 053804.}

\bibitem{Hentschel2010}
A. Hentschel and B. C. Sanders, 
Machine Learning for Precise Quantum Measurement, 
\href{https://doi.org/10.1103/PhysRevLett.104.063603}
{Phys. Rev. Lett. \textbf{104} (6) (2010) 063603.}

\bibitem{Hentschel2011}
A. Hentschel and B. C. Sanders, 
Efficient Algorithm for Optimizing Adaptive Quantum Metrology Processes, 
\href{https://doi.org/10.1103/PhysRevLett.107.233601}
{Phys. Rev. Lett. \textbf{107} (23) (2011) 233601.}

\bibitem{Lovett2013} 
N. B. Lovett, C. Crosnier, M. Perarnau-Llobet, et al.,  
Differential Evolution for Many-Particle Adaptive Quantum Metrology, 
\href{https://doi.org/10.1103/PhysRevLett.110.220501}
{Phys. Rev. Lett. \textbf{110} (22) (2013) 220501.}

\bibitem{Rambhatla2020}
K. Rambhatla, S. E. D’Aurelio, M. Valeri, et al., 
Adaptive phase estimation through a genetic algorithm, 
\href{https://doi.org/10.1103/PhysRevResearch.2.033078}
{Phys. Rev. Res. \textbf{2} (3) (2020) 033078.}

\bibitem{Berni2015}
A. A. Berni, T. Gehring, B. M. Nielsen, et al.,  
\textit{Ab initio} quantum-enhanced optical phase estimation using real-time feedback control, 
\href{https://doi.org/10.1038/nphoton.2015.139}
{Nat. Photon. \textbf{9} (9) (2015) 577.}

\bibitem{Tsitouras2011}
C. Tsitouras, 
Runge-kutta pairs of order 5 (4) satisfying only the frst column simplifying assumption,
\href{https://doi.org/10.1016/j.camwa.2011.06.002}
{Comput. Math. Appl. \textbf{62} (2) (2011) 770.}

\bibitem{gitlink}
\href{https://github.com/QuanEstimation/QuanEstimation.jl}
{\small{github.com/QuanEstimation/QuanEstimation.jl}}

\bibitem{doclink}
\href{https://quanestimation.github.io/QuanEstimation/}
{\small{quanestimation.github.io/QuanEstimation/}}

\end{thebibliography}
\end{document}